# Undisturbed Mesosphere Optical Properties from Wide-Angle Frequent Twilight Sky Polarimetry

Oleg S. Ugolnikov[1], Igor A. Maslov[1,2]

[1]Space Research Institute, Russian Academy of Sciences, Russia
[2]Sternberg Astronomical Institute, Moscow State University, Russia

Corresponding author e-mail: ougolnikov@gmail.com

The paper describes the first results of all-sky polarization measurements of twilight background started in central Russia in the very beginning of summer 2011. Time-frequent data of sky intensity and polarization over the wide range of sky point zenith distances is used to separate the multiple scattering and to build the altitude dependency of scattering coefficient and polarization in the mesosphere (altitudes from 60 to 90 km) at different angles. Undisturbed structure of the mesosphere without the noticeable aerosol stratification during the observation days allows to estimate the atmosphere temperature at these altitudes.

## 1. Introduction

Twilight sounding of the Earth's atmosphere is well-known tool of its optical structure investigations. Being in use for more than the century, it became an object of first quantitative theory in 1923 [1]. Until the start of lidar, plain and satellite measurements, the method (along with the meteors moderation and burning observations) was the only way to investigate high layers of the atmosphere.

In present time this method remains effective for some atmospheric layers, first of all, the mesosphere. It is so high for plains and for most part of lidars, but it is low and dense for the satellite flights. It is also very transparent for the space remote absorption sounding (the method successfully works for troposphere and stratosphere). The basic way of mesosphere optical investigations is the measurements of scattered radiation.

Mesosphere is the object of interest since it is the bounder between lower dense parts of the atmosphere and outer parts with principally different physical properties. The cosmic dust particles falling to the atmosphere are moderated in the mesosphere (small particles waste the kinetic energy without burning). The number of these particles increases during the maxima of major meteor showers [2]. Mesosphere is also the coldest atmosphere layer. Summer temperature values in mid-latitudes can fall down to 180 K. In these conditions the polar mesospheric (or noctilucent) clouds, highest in the atmosphere, can appear. The role of condensation nuclei for these clouds can be played by the cosmic dust particles.

The basic advantage of twilight method of atmosphere sounding is related with the possibility of simultaneous measurements in different sky points during the whole twilight period. This allows to find the scattering function of atmospheric volume unit in a wide range of scattering angles – that is difficult for the lidar or space investigations. The twilight growth of the Earth's shadow above the observer (that can be quite fast compared with the time of atmospheric properties variation) allows to cover wide range of altitudes.

The possibilities of twilight method for the mesosphere were discussed for many years [3]. The basic factor restricting the altitude range is the multiple scattering of solar radiation in the atmosphere, especially in the lower layers. According to optimistic estimations ([3] and references therein), single scattering contribution would remain noticeable during the whole twilight period, expanding the work altitude range of twilight method up to the thermosphere. However, analysis of observational effects (starting from [4]) and more precise numerical simulation that had become possible just recently [5] had restricted this range by the mesopause.

Polarization measurements of the twilight sky background sufficiently improve both the accuracy of multiple scattering separation and the information about scattering properties of the different atmosphere layers. The base of polarization method was established in [6]. Polarization data were used to find the multiple scattering contribution during the light twilight period [7, 8], it was found out to be more than by previous estimations [3] but in good agreement with precise numerical calculations [9]. The behavior of the sky polarization value during the twilight clearly reflects the contribution changes between single (molecular and aerosol) and multiple scattering [10].

Using the polarization twilight sky measurements near the zenith, the dust scattering admixture was found in stratosphere [11] and mesosphere [10], that was related with Rabaul volcano eruption in 2006 and strong Leonids meteor shower maximum in 2002, respectively. But small camera field size (about 8°) did not allow to investigate these scattering components numerically and to build their scattering matrices. In this paper we will discuss the first results of wide-field polarization twilight sky measurements, focusing on dark twilight period and highest atmosphere layer that can be investigated by such a way – the upper mesosphere.

## 2. Observations.

The wide-angle polarization measurements had started in central Russia (55.2°N, 37.5°E), 60 km southwards from Moscow. The measuring device consists of “fish-eye” lens with two transitive lenses and Sony DSI-Pro CCD-matrix. The rotating polarization filter is installed near the transitive lenses, the light rays cross it by quite small angles to the axis. The Meade G and IR-blocking filters restrict the 100 nm-wide spectral range with effective wavelength equal to 540 nm. The field size of the camera is about 140°. The camera axis is directed towards the zenith, its exact position is determined by the stars images on dark twilight and night frames. The measurements start during the daytime (before the sunset), continue through all the night and finish after the sunrise. The exposure time varies from 0.001 second (the light twilight) to 15 seconds (the dark twilight and the night). The imaging frequency is about 1 frame per 2 seconds during the light twilight and 1 frame per 18 seconds during the night. The time interval between polarization filter moves by 120° is 20 seconds in the light twilight and 2 minutes in the night. So, the series contain a number of frames, and the interpolation interval for polarization calculation is less than the time of sufficient change of sky brightness.

This paper is based on the data of two consecutive nights of May, 31 and June, 1, 2011. These nights were characterized by stable good weather and atmosphere transparency. Being close to New Moon (June, 1), these nights were also free from scattered moonlight. The early summer period is especially interesting since it is the time of strong mid-latitudes mesosphere cooling prior to noctilucent clouds appearance. The mid-latitude twilight in this season is long and never turns to the deep night, but it contains the long dark twilight and nightfall periods (solar zenith angle $z_0$ more than 100°) in the middle. This increases the data density and method accuracy.

Figure 1 shows the dependencies of sky background intensity in the zenith on the Universal Time (UT) during the both nights. The curves almost coincide with each other, but the one for the June, 1 shows the decrease near the local midnight caused by the partial solar eclipse that occurred at the Earth’s terminator northwards from the observation point and weakened the Sun emission by about 2 times. Since the eclipse was not close to total and occurred in the dark twilight period with multiple scattering domination, it decreased the brightness of all sky points by the same value and had no influence on the mesosphere parameters obtained below.

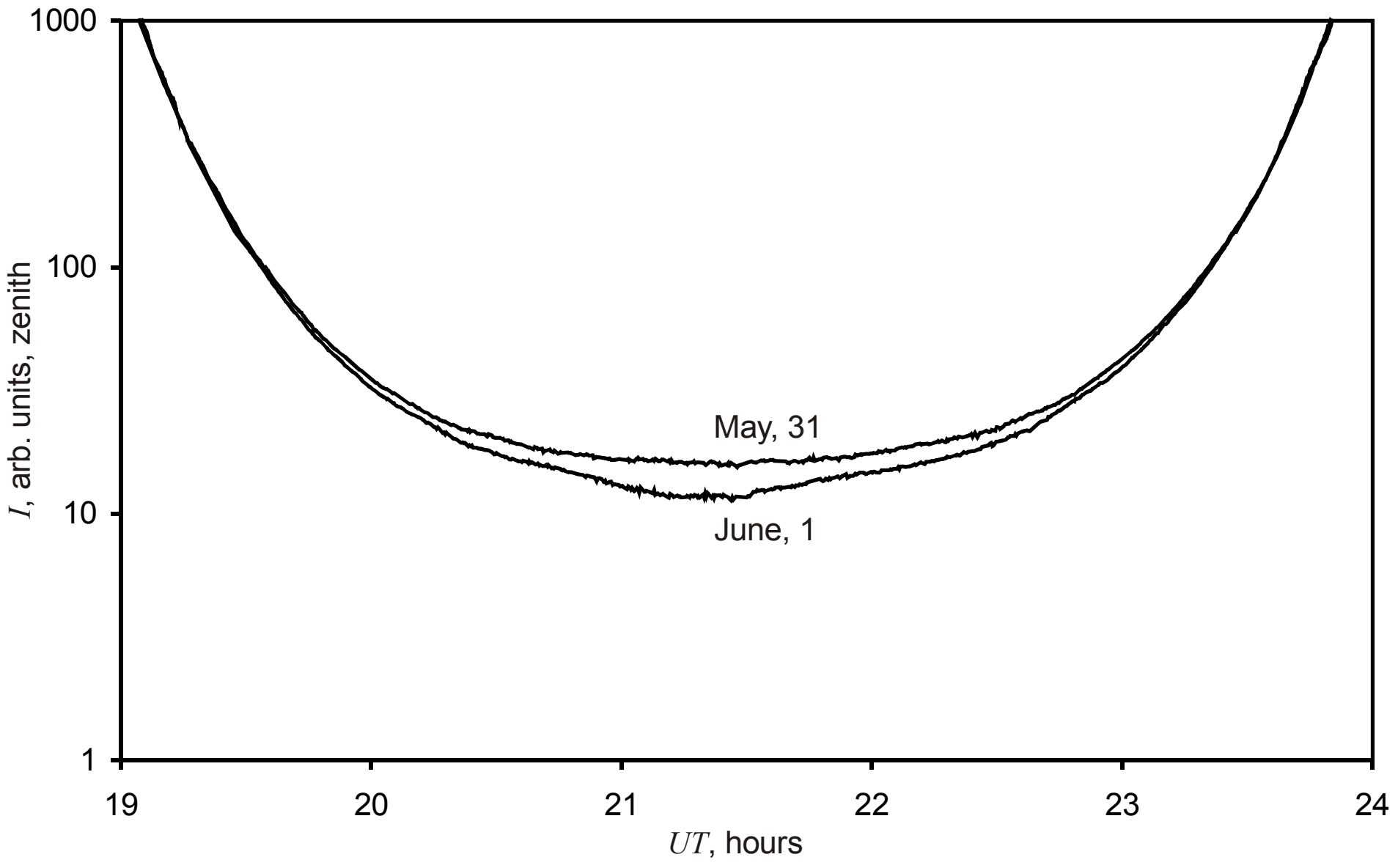

*Figure 1. Twilight sky background intensity in the zenith depending on time during two observational nights. The effect of partial solar eclipse is seen near the local midnight of June, 1.*

The geometry of single scattering during the twilight is shown in the Figure 2. The solar zenith angle $z_0$ is more than 90° and lower layers of the atmosphere above the observer (point O) are immersed into the Earth's shadow. Moreover, solar radiation transferring near the shadow border through the low dense atmosphere layers is sufficiently absorbed. Effective scattering occurs along the path elevated above the shadow by the altitude $H$ (the "twilight ray" model [3]). During the dark twilight the value of $H$ for 540 nm is equal to 18 km (with account of atmospheric temperature distribution and Chappuis ozone absorption with total ozone content equal to 300 Dobson units).

Such value of $H$ allows to disregard the refraction, since the maximal refraction angle ρ (for the ray crossing the lower atmosphere and reaching the mesosphere) is about 6′ or $2 \cdot 10^{-3}$ radians, that is just 40% of the visible radius of the solar disk. The decrease of scattering point altitude in the dusk sky area will not be more than $R(z_0-\pi/2)\rho$ or 2 km ($R$ is the Earth's radius), that is less than the vertical resolution of the twilight sounding method for the mesosphere.

Let us denote the zenith distance of observational point by $z$. Let this point be at the solar vertical, and value of $z$ is positive in the dusk (dawn) area and negative in the opposite part of the sky. When the Sun is close to horizon, the twilight ray is almost horizontal, and effective scattering altitude $h(z,z_0)$ does not rapidly depend on $z$, but during the darker stages of twilight this dependency becomes stronger (see Figure 2), leading to sky brightness excess in the dusk area with positive $z$.

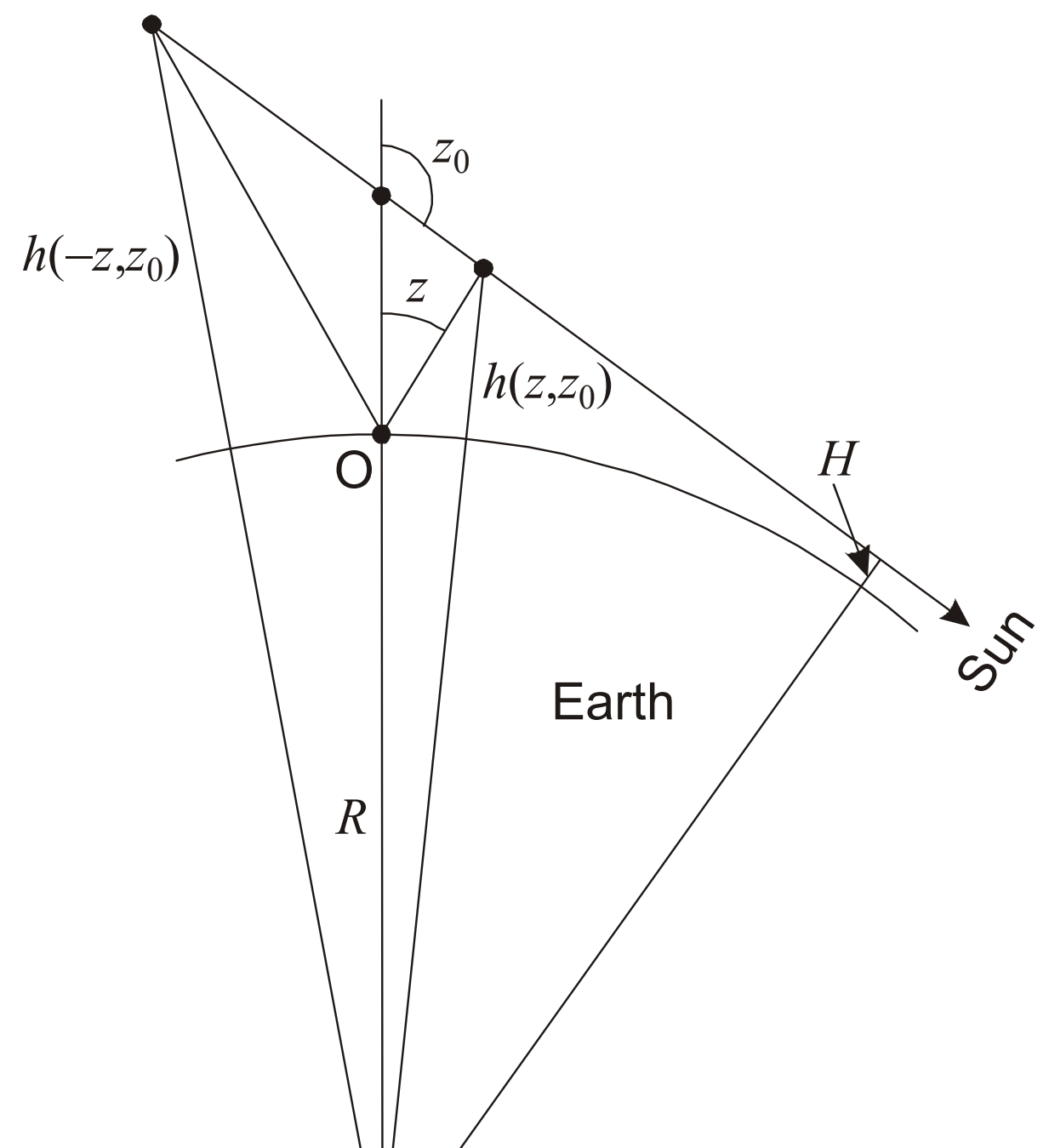

*Figure 2. Geometry of single scattering during the twilight.*

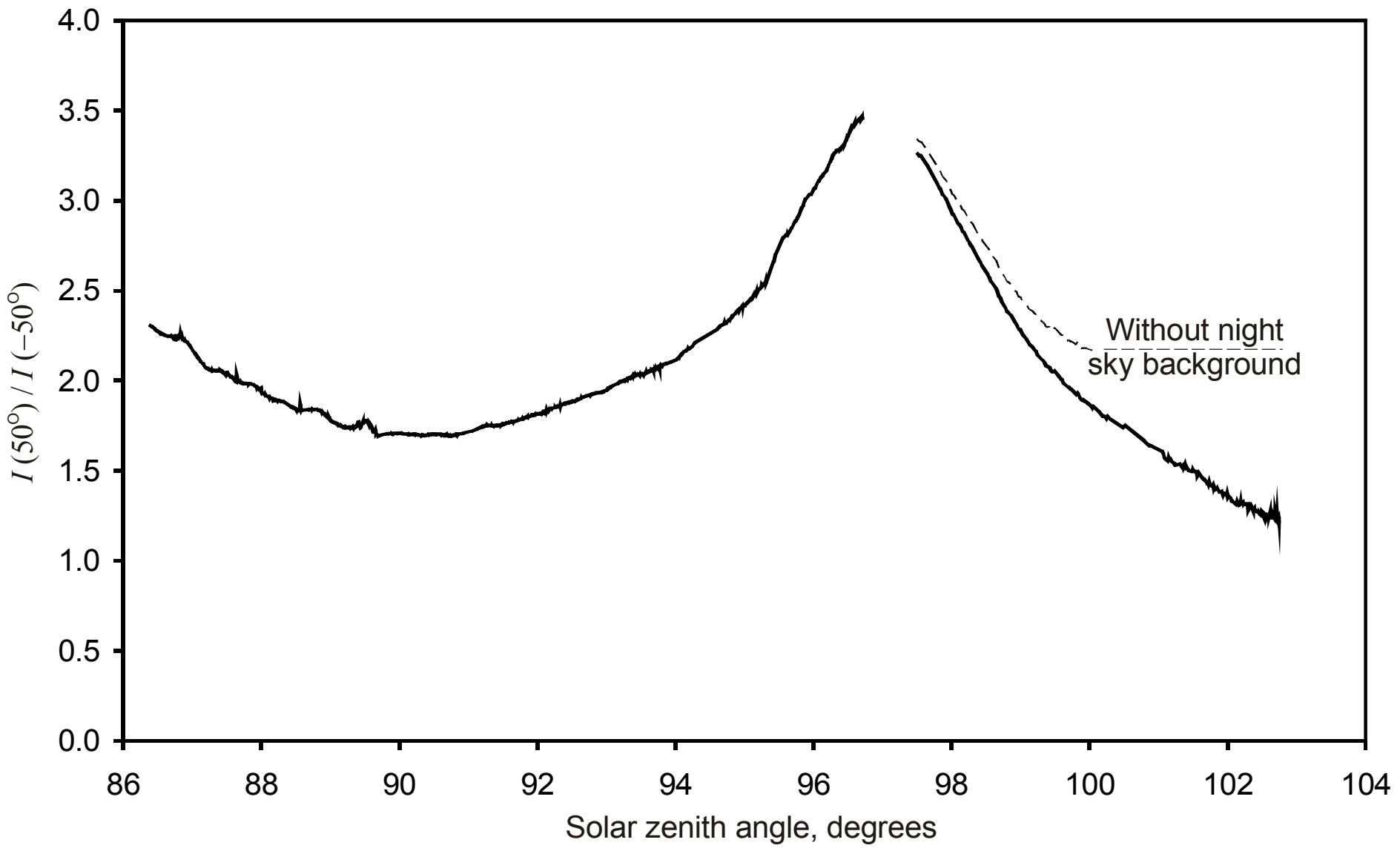

*Figure 3. Intensity ratio of the symmetrical solar vertical points at the evening twilight of May, 31.*

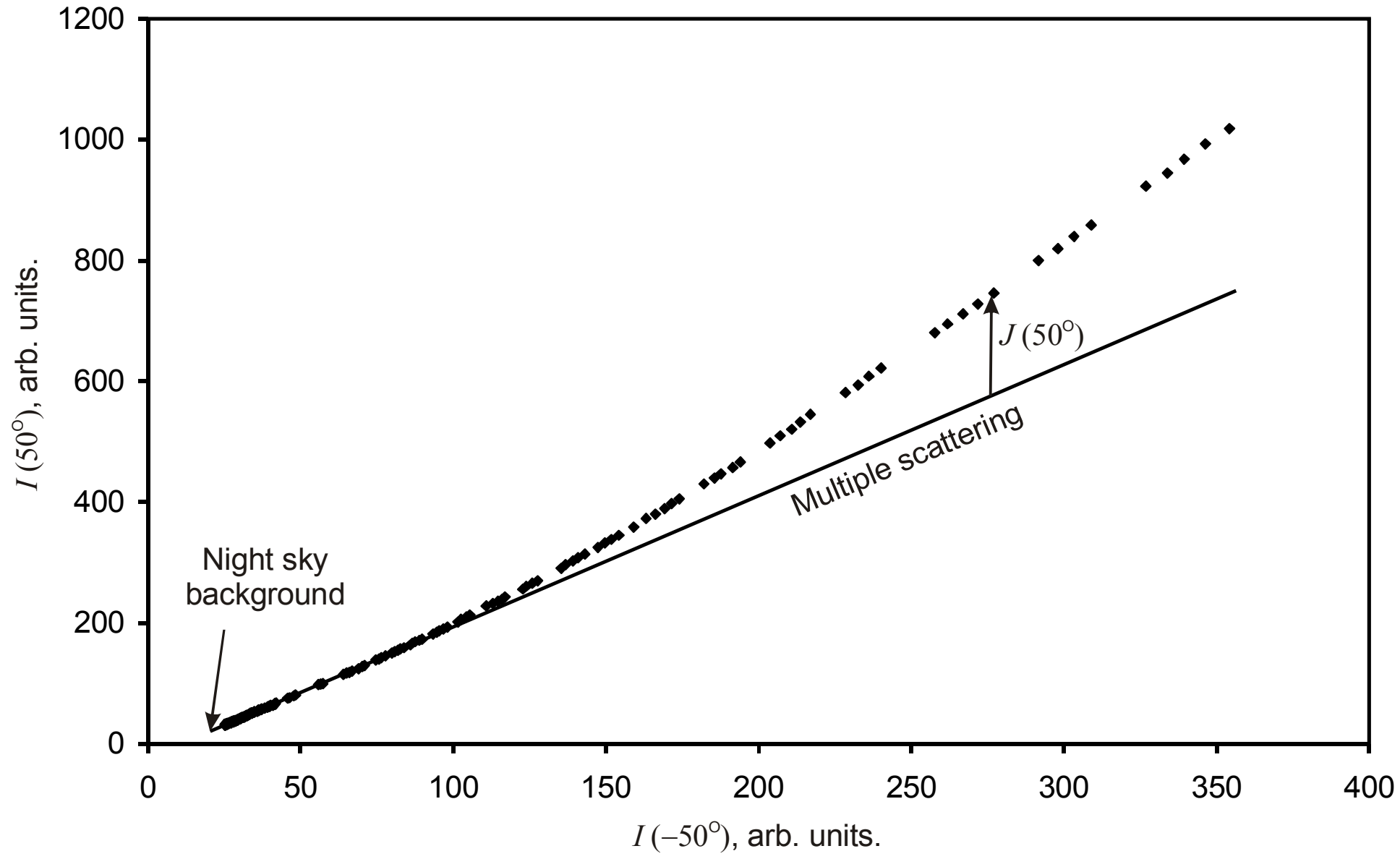

*Figure 4. Comparison of intensity values in the symmetrical solar vertical points (the same twilight).*

This effect can be seen in the Figure 3, where the dependency of sky intensity ratio in the symmetrical solar vertical points $I(z)/I(-z)$ on the solar zenith angle is shown for the evening of May, 31, $z = 50°$. The ratio does increase until solar zenith angle about 97° but then decreases. The same effect was observed earlier [7] and cannot be related with single scattering (even if we had assumed this, than the scattering properties of high atmosphere would not depend on the altitude and the sky brightness would not decrease with further Sun depression). This twilight stage the multiple scattering starts to dominate. At solar zenith angles more than 100° the ratio in the Figure 3 decreases by linear law and turns to constant if we estimate and subtract night sky background intensity that becomes noticeable during this stage of twilight.

Constant intensity ratio in the symmetrical solar vertical points is the property of multiple scattering used in [6, 12]. It is also seen in the Figure 4, where the intensity values are compared one with another. Multiple scattering dominates at both zenith distances $z$ and $-z$ ($z$=50° in the figure) when the solar zenith angle is more than 100°, the corresponding points lay along the straight line shown in the figure. When the solar zenith angle is a little bit less, the intensity $I(-z, z_0)$ is still consists of multiple scattering, but dusk area background $I(z, z_0)$ is also contributed by single scattering component $J(z, z_0)$. The corresponding dots are deviated from the line in the Figure 4, and this way the single scattering component $J(z, z_0)$ can be determined. It can be done until the single scattering contribution appears also at the zenith distance $–z$. This border will be defined below. We should note that the intensity comparison is made only for the symmetrical solar vertical points and possible radial flat-field uncertainties of wide-angle camera with "fish-eye" lens do not affect the results. In several papers ([12] for example) the procedure is continued to the less solar zenith angles using the linear dependency of multiple scattering intensity logarithm on the solar zenith angle, but such extrapolation can lead to sufficient errors in results [7].

Figure 5 shows the dependency of sky background polarization $P_0$ on the solar zenith angle for the zenith and solar vertical points with $z$ equal to ±50°. Data correspond to the evening twilight of June, 1, polarization means the Stockes vector component ratio $I_2/I_1$, positive for the polarization direction perpendicular to the scattering plane and negative for the direction parallel to the scattering plane. Zenith curve reveals all the twilight stages described in [10] but with slight variations of borders: early twilight ($z_0$ less than 93°), where the polarization increases with $z_0$ due to the immersion of tropospheric aerosol into the Earth's shadow; light twilight ($z_0$ from 93° to 95°) with maximal polarization and almost constant ratio of single and multiple scattering; transitive twilight ($z_0$ from 95° to 99°), when the polarization falls due to decrease of single scattering contribution; dark twilight ($z_0$ from 99° to 100°) totally dominated by multiple scattering with constant polarization and, finally, nightfall ($z_0$ more than 100°) when polarization decreases due to the night sky background contribution.

The picture differs far from zenith, at $z$ equal to ±50°. During the light stages of twilight polarization increases in the dusk area and decreases (with almost the same gradient) in the opposite sky part. This effect is related with the motion of maximal polarization point of single scattering (about 90° from the Sun) following the Sun. It became the basis of the single scattering separation method [7, 8] for these twilight stages. After that the single scattering contribution and the polarization value decrease, but dark twilight comes at different solar zenith angles: about 97.5° for $z$=–50° and 100° for $z$=50° (arrows in the Figure 5). These are the borders of interval, where the single scattering at $z$=50° can be separated by the method described in this paper (however, good accuracy is reached only for $z_0$<99.3°). The same procedure can be hold for other values of $z$, but with other border solar zenith angles.

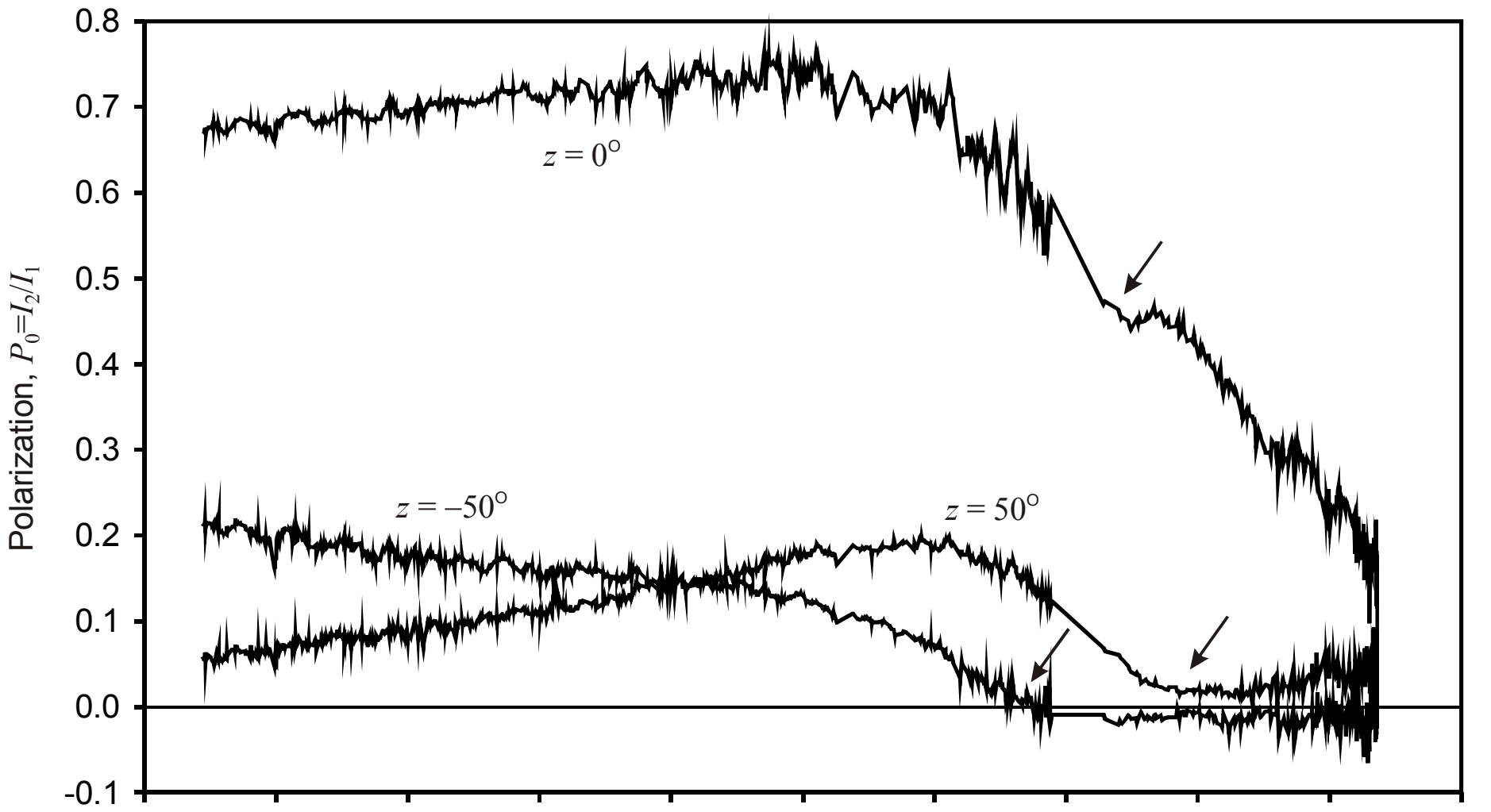

*Figure 5. Twilight sky background polarization depending on solar zenith angle (evening of June, 1). Arrows show the beginning of dark twilight stage.*

Dark twilight polarization values for $z$ equal to ±50° are almost constant and the same, it is the property of multiple scattering polarization in the symmetric solar vertical points [7]. But small difference and variations with solar zenith angle are seen. They can be related with night sky background and its self polarization, increased by the light echo from Moscow in the north. This problem is taken into account in the procedure of single scattering separation described below.

**3. Single scattering separation procedure.**

Twilight sky background intensity $I(z, z_0)$ is the sum of single scattering component $J(z, z_0)$, multiple scattering component $j(z, z_0)$ and night sky background $n(z)$. Single scattering is noticeable if solar zenith angle $z_0$ is less than limit value $z_{0S}(z)$. We consider the symmetric solar vertical points with fixed zenith distances $z$ and $-z$ and solar zenith angles interval:

$$z_{0S}(-z) < z_0 < z_{0S}(z). \quad (1)$$

The twilight sky intensity values are:

$$I(z,z_0) = J(z,z_0) + j(z,z_0) + n(z);$$
$$I(-z,z_0) = j(-z,z_0) + n(-z). \quad (2)$$

Basing on data from neighbor interval $z_0 > z_{0S}(z)$ where $J=0$, we know:

$$j(z,z_0) + n(z) = A(z) + B(z)\cdot(j(-z,z_0) + n(-z)). \quad (3)$$

This is expressed by a straight line in the Figure 4. Coefficients $A$ and $B$ do not depend on $z_0$, so this equation can be rewritten as

$$j(z,z_0) = B(z)\cdot j(-z,z_0);$$
$$n(z) = A(z) + B(z)\cdot n(-z). \quad (4)$$

Calculating the values of $A$ and $B$ from the diagram in the Figure 4, we put them into equation (2) and write the expression of single scattering intensity:

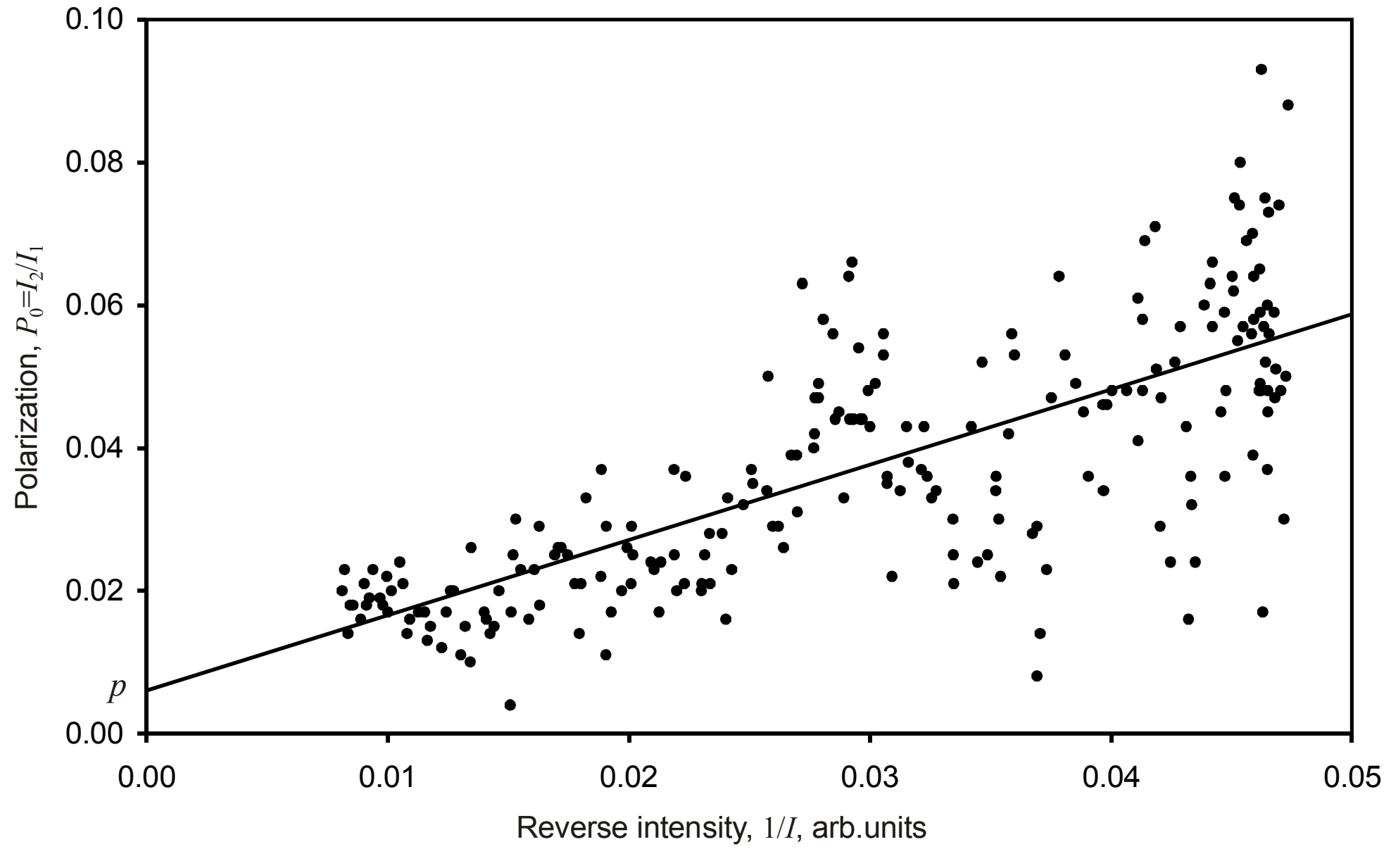

*Figure 6. Correlation of reverse intensity and polarization for dark twilight stage of evening, June 1, z=50°.*

$$J(z, z_0) = I(z, z_0) - A(z) - B(z) \cdot I(-z, z_0). \tag{5}$$

Here we should note that the solar eclipse of June, 1 occurred during the dark twilight when all $J$ values were equal to zero and $j$ values were decreased by the same amount not changing the parameter $B$.

Polarization values of multiple scattering ($p$) and night sky background ($q$) can be assumed to be independent on $z_0$ [7]. For positive values of $z$ during the dark twilight ($z_0 > z_{0S}(z)$) the sky polarization is equal to

$$p_D(z, z_0) = \frac{j(z, z_0)p(z) + n(z)q(z)}{j(z, z_0) + n(z)} = p(z) + \frac{(q(z) - p(z))n(z)}{I(z, z_0)} = p(z) + \frac{C(z)}{I(z, z_0)}. \tag{6}$$

Here $p_D(z, z_0)$ is the polarization of multiple scattering together with night sky background. Having put the experimental values of polarization ($p_D$) and reverse intensity ($1/I$) for $z_0 > z_{0S}(z)$ on the diagram (shown in the Figure 6 for evening of June, 1), we find the values of $p$ (see Figure 6) and $C$. If we take another $z_0$ interval (1), than the intensity $I$ will not be equal to sum of $j$ and $n$, and the sky background polarization $P_0$ will not be equal to $p_D$ owing to single scattering contribution:

$$p_D(z, z_0) = p(z) + \frac{C(z)}{j(z, z_0) + n(z)} = p(z) + \frac{C(z)}{A(z) + B(z)I(-z, z_0)}; \tag{7}$$

$$P_0(z, z_0) = \frac{J(z, z_0)P(z, z_0) + ((j(z, z_0) + n(z))p_D(z, z_0))}{J(z, z_0) + j(z, z_0) + n(z)} =$$

$$= \frac{J(z, z_0)P(z, z_0) + p(z)(A(z) + B(z)I(-z, z_0)) + C(z)}{I(z, z_0)}. \tag{8}$$

Finally, the polarization of single scattering component (normalized second component of Stockes vector):

$$P(z, z_0) = \frac{I(z, z_0)P_0(z, z_0) - p(z)(A(z) + B(z)I(-z, z_0)) - C(z)}{J(z, z_0)}. \tag{9}$$

Formulae (5) and (9) give us the required values of intensity and polarization of single scattered radiation.

**4. Results and conclusion.**

Figure 7 shows the different twilights polarization dependencies of single scattering $P(z_0)$ for two $z$ values (40° and 50°) on the effective altitude of single scattering $h(z, z_0)$. The results are compared with the Rayleigh scattering polarization (with depolarization parameter 0.06) shown by the lines. The last value is slowly changing with the scattering angle equal to ($z_0 - z$). We see the picture usual for the undisturbed mesosphere – polarization does not sufficiently change (just slightly decreases) over the altitude range and its value is a little bit less than the Rayleigh one. Taking into account that aerosol scattering polarization sufficiently weaker than molecular scattering one, we can estimate the aerosol scattering intensity as 20-30% of Rayleigh scattering with little increase with altitude. This result is in good agreement with space ultraviolet observations [13].

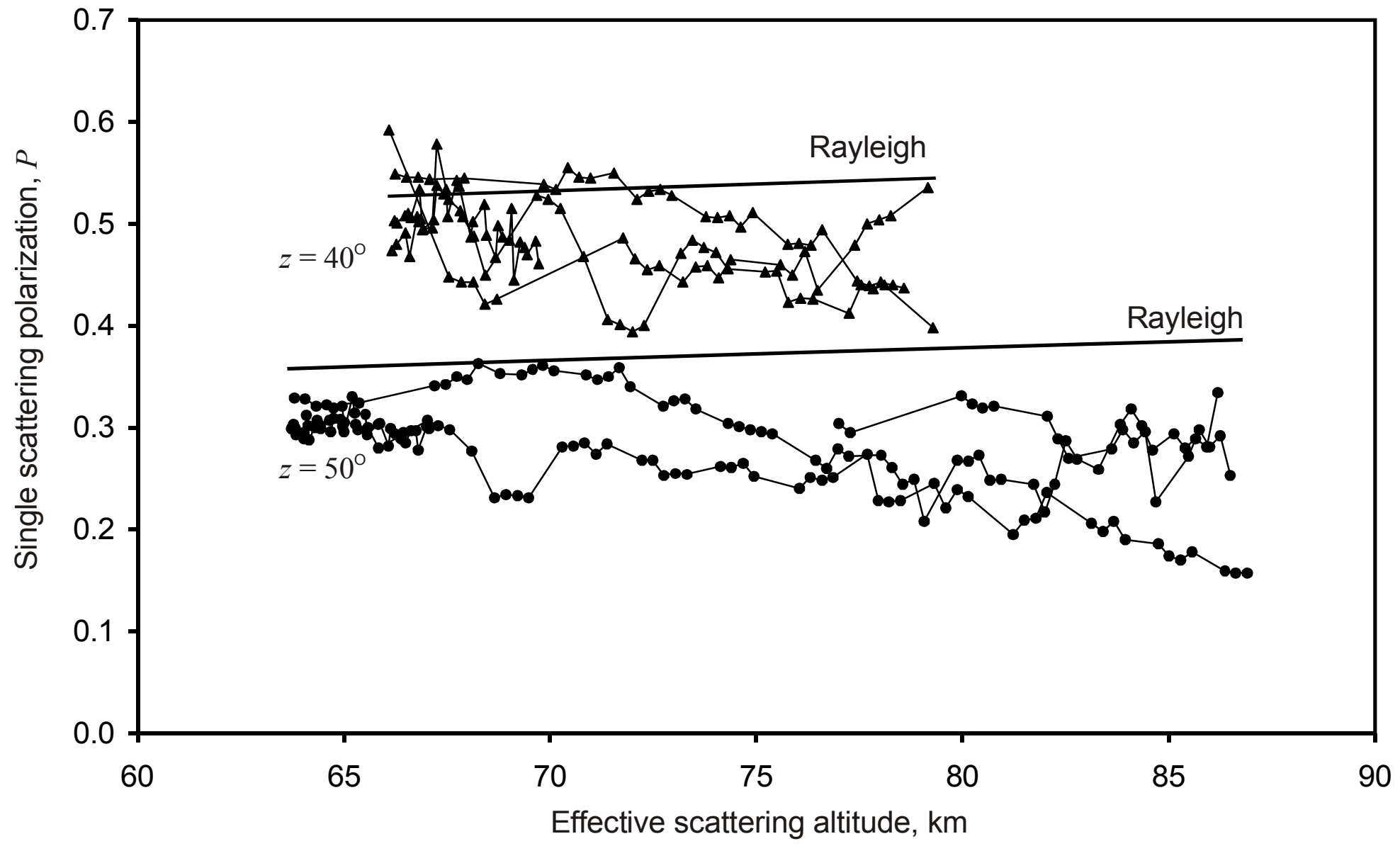

*Figure 7. Altitude dependencies of single scattering polarization for zenith distances 40° and 50° compared with the Rayleigh scattering values.*

The single scattering intensity dependencies $J(z_0)$ on the effective scattering altitude are shown in the Figure 8 for the zenith distance 50°. The corresponding $z_0$ values for this $z$ are also noted along the x-axis. For transitive and dark twilight stages ($z_0$>95°) scattering points are above the dense atmosphere layers, the "twilight ray" optical depth values before and after scattering does not depend on $z_0$ and the intensity is proportional to the scattering coefficient at the effective scattering altitude by the angle $(z_0 - z)$. Taking into account the slow change of this angle and of aerosol scattering contribution, it is expected that the value $J(z_0)$ would be proportional to the atmosphere density at the altitude $h(z_0)$ for the fixed $z$ value. Figure 8 shows that it is, the intensity logarithm dependencies are close to the linear, and temperature estimated from them (in particular, the mesopause temperature) coincides with the typical values for mid-latitude early summer. This can be used for temperature measurements of undisturbed mesosphere. The temperature exactness can be improved if we take the refraction into account and use the "twilight layer" integration instead of "twilight ray" model. But it is not necessary for aerosol scattering investigations.

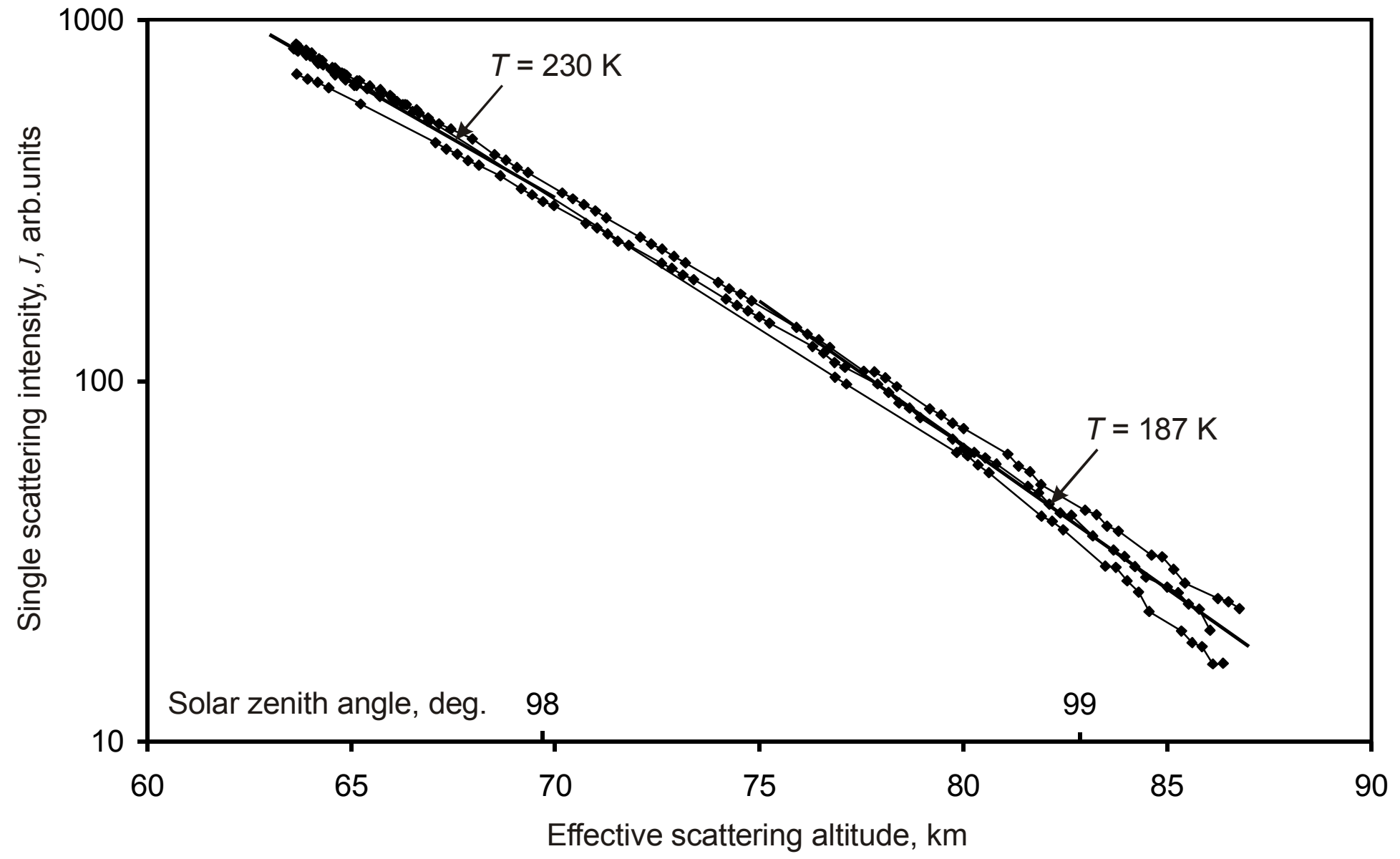

*Figure 8. Altitude dependencies of single scattering intensity for zenith distance 50°.*

The accuracy of single scattering intensity and polarization measurements seems to be enough not only to fix, but to measure the increased aerosol concentration in the mesosphere and investigate its evolution in time. This can take place after the meteor showers maxima, when the upper mesosphere is bombarded by meteor dust particles. The polarization change can be registered even in the zenith ([10] and references therein), and the increased scattering in the dusk area will be visible at higher solar zenith angle $z_0$ (for the same altitude $h(z, z_0)$) on the weaker background of multiple scattering. Such observations are especially interesting in mid-latitude summer, when strong mesosphere cooling can lead to the appearance of noctilucent clouds, and the twilight duration and maximal $z_0$ value are optimal to hold these measurements.